# Teaching Classical Mechanics using Smartphones

Joel Chevrier[1], Laya Madani[2], Simon Ledenmat[2], Ahmad Bsiesy[2]

[1] *Université Joseph Fourier - BP 53 38041 Grenoble cedex 9*

[2] *CIME Nanotech, Grenoble INP/UJF, BP 257, 38016 Grenoble Cedex 1*

Classical Mechanics is taught everywhere in the world [1]. It is usually a first year course at university although some countries introduce it before. No surprise as it is very much related to our daily experience (displacement, sport, transport,...) and to spectacular human achievements (plane, satellite, ...). Based on the classical description of space and time, this course introduces descriptions of trajectories and movements and their relationships with forces through Newton's laws. It then uses vectors to represent frames, acceleration and velocity. It describes energy and the associated conservation law.

Following articles published in the Physics Teacher which illustrates this [2,3], we have looked at a smartphone with Classical Mechanics in mind. Steve Jobs presented the iPhone as «perfect for gaming» [4]. Thanks to their microsensors connected in real time to the numerical world, physics teachers could add that smartphones used nowadays by so many students in daily life, are «perfect for teaching science». The central point of our work is the software iMecaProf [4] that displays in real time the measured data on a screen. The visual representation is built upon the formalism of Classical Mechanics. iMecaProf receives data 100 times a second, from iPhone sensors through a WIFI connection using so far the application Sensor Data [5]. Data are the three components of the acceleration vector in the smartphone frame and smartphone orientation through 3 angles (yawl, pitch and roll). For a circular movement (uniform or not, pendulum), iMecaProf uses independent measurements of the rotation angle θ, the angular speed $d\theta/dt$ and the angular acceleration $d^2\theta/dt^2$.

We have systematically identified where iMecaProf can experimentally accompany a course on Classical Mechanics. Different levels in iMecaProf have then been defined which define a pedagogical progression. **Level 0** is somewhat reminiscent of Steve Jobs show [4]: a virtual representation of the real smartphone on a visual display with no added scientific tool. Manipulation of the real smarphone controls the virtual one orientation. Only changes in 3D orientation are reproduced on the screen and not the translations. One can directly visualize how the moon rotates and the existence of its dark side. **Level 1** adds frames and unit vectors to level 0. Fixed unit vectors are associated to the lab frame (O,**X,Y,Z**) and rotating unit vectors to the smartphone frame (O,**x$_{iPh}$,y$_{iPh}$,z$_{iPh}$**). **Level 2** splits the screen in four parts as shown in figure 1: upper left keeps image of level 1 whereas three other parts show projections of unit vectors attached to the smartphone. Users can practice orientation of real smartphone to obtain different projections of unit vectors **x$_{iPh}$, y$_{iPh}$** and **z$_{iPh}$** in planes OXY, OXZ and OYZ. One can try to have the three projections of unit vectors **x$_{iPh}$, y$_{iPh}$, z$_{iPh}$** equal in all three planes OXY, OXZ and OYZ. A game associated to this level proposes the user to control orientation of the real smartphone so that its artifact is always aligned with a second virtual smartphone whose orientation is controlled on screen by the computer. In **Level 3,** iMecaProf enters dynamics: acceleration **a** and velocity **v** vectors are shown on screen and in real time, in lab frame and in smartphone frame. Level 3 options are designed to specifically offer a detailed study of cases most often studied in Basics Mechanics. In 2D option, movements are rotation: circular motion (uniform or not) and pendulum. In 1D option, displacements include free fall or elevator movement, and 1D oscillation to experimentally explore mechanical resonance. The screen on visual display is then split in 4 parts as shown in figure 2. Upper left part is again level 1 image. Upper right shows in lab frame the unit vectors of both frames (lab and smartphone), acceleration **a** and velocity **v**. Lower right shows in rotating smartphone frame, the associated unit vectors, plus acceleration **a** and velocity **v**. Lower left is used to plot versus real time, the potential energy, the kinetic energy and their sum, the mechanical energy. In the 1D case, acceleration and velocity vectors are displayed real time as arrows. Their values are plotted real time in a single graph. A second game has been here introduced so that user can play with acceleration. The computer defines a 1D oscillating movement with period from 0,5 sec to 2 sec and amplitude between 10 cm and 50 cm. Authors reasonably managed to control smartphone movement by hand to reproduce period and amplitude. We hope users will achieve better results as far as phase is concerned.

Levels 0,1 and 2 require no specific material or setup to use iMecaProf as above described. The smartphone is in user hand. The minimum set up used in level 3 is very limited. Circular motion can be studied using a swivel chair, pendulum using the smartphone charge cable and oscillation by putting the smartphone into a plastic bag attached to an elastic rubber. Insertion of the smartphone in a more sophisticated setup can be easily considered as long as smartphone size is not an issue and if its movement is not too fast.

In conclusion, motivations to develop iMecaProf [6, 7] are : i) using a personal computer and a smartphone, it provides a complete teaching environment for practicals associated to a Classical Mechanics course. This is based on a visual, real time and interactive representation of measured data using the formalism of Classical Mechanics; ii) using smartphones is more than using a set of sensors. iMecaProf shows students that important concepts of physics they here learn, are necessary to control daily life smartphone operations; iii) this is practical introduction to mechanical microsensors that are nowadays a key technology in advanced trajectory control.

First version of iMecaProf [6, 7] is now ready to be used and will tested this academic year in Université Joseph Fourier (Grenoble, France)

[5] http://wavefrontlabs.com/Wavefront_Labs/Sensor_Data.html

[6] iMecaProf can be downloaded at:

https://dl.dropbox.com/s/7qln80wiqbxl9ne/iMecaProf%203.zip?dl=1

[7] http://www.youtube.com/watch?v=r4_3Fv9GKQ4

**Figures :**

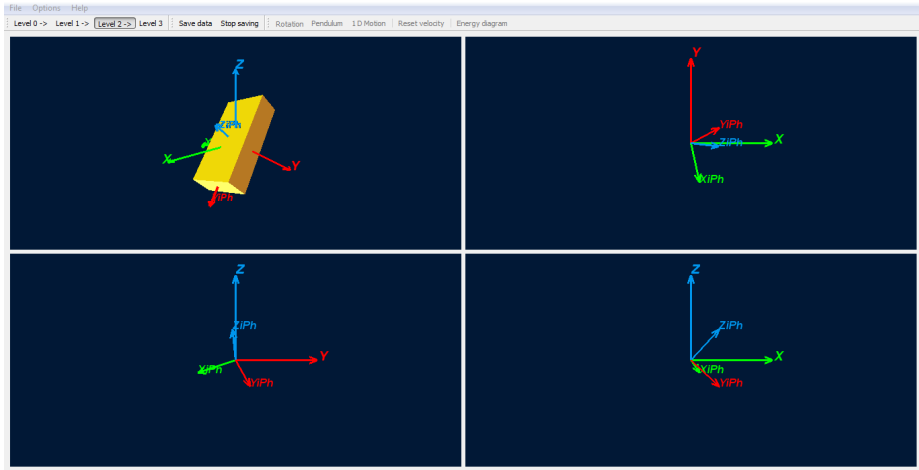

Figure 1: Upper left corner keeps image of level 1 (see text) whereas three other parts show projections of unit vectors attached to the smartphone (**x$_{iPh}$, y$_{iPh}$, z$_{iPh}$**) in OXY, OZY and OXZ planes defined in the lab frame.

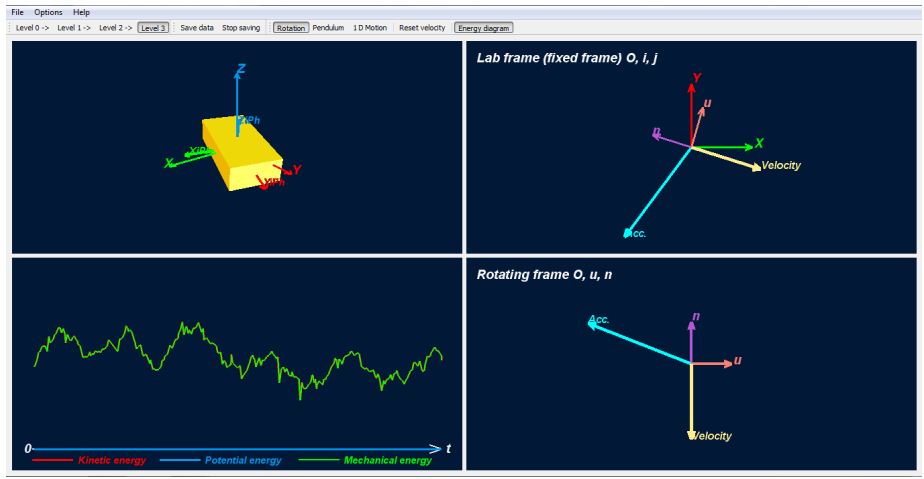

Figure 2: iPhone on a rotating swivel chair. Non uniform circular rotation. Upper left part is again level 1 image (see text). Upper right shows in lab frame the unit vectors of both frames (polar vectors **u** and **n** here used in the smartphone frame are rotating), acceleration **a** and velocity **v**. Lower right shows in rotating smartphone frame, the associated unit vectors **u** and **n** (now immobile), plus acceleration **a** and velocity **v**. Lower left is used to plot versus real time, the potential energy (here equal to zero), the kinetic energy and their sum, the mechanical energy.